# Experimental Determination of an Extended DC Servo-Motor State Space Model: An Undergraduate Experiment

**Ahmad A. Masoud (a, corresponding author) 1, Mohammad Abu-Ali 2, Ali Al-Shaikhi 1**

[1] Electrical Engineering Department, KFUPM, Dhahran Saudi Arabia
[2] Electrical and Computer Engineering, Technical University of Munich, Munich, Germany

Corresponding author: Ahmad A. Masoud (e-mail: masoud@kfupm.edu.sa).

**ABSTRACT** State space systems and experimental system identification are essential components of control education. Early introduction of these tools to the curriculum of a control laboratory in an interconnected and accessible manner that does not over-dilute the concepts is important in two ways. First, it facilitates a student's transition to more senior and graduate level control concepts. It also provides an effective link to industrial applications. This paper provides two novel experimental procedures for directly identifying the state space model of a DC motor in an undergraduate control laboratory. The procedures do not require any specialized hardware and can be performed using standard laboratory equipment. They also do not place any simplifying assumptions on the motor's model. The first procedure is based on direct pseudo inversion of the state space model. It does not require advanced knowledge of the state space approach or signal filtering. It is easy to understand and suits a first control laboratory. The second procedure is more efficient. It is based on the Markov approach commonly used for realizing, indirectly, a state space model from an estimated transfer function. While the procedure is designed for use by undergraduate students, it requires relatively advanced knowledge in state space that makes it suitable for a second undergraduate control laboratory.

**INDEX TERMS** Control Education, Control Systems, Motor Modeling, System Identification,

## I. INTRODUCTION

The importance of electrical motors to both academia and industry [1] is obvious. DC motors are usually adopted as the servo-process of choice in control laboratories. Investigating the control of motors is usually preceded by an experiment to determine their transfer function. A common practice in a 1st control laboratory is to model the motor using its step response assuming a first order approximate velocity transfer function. The justification is that the electrical time constant of the motor is negligible compared to its mechanical time constant. This assumption, as shown in the paper, creates nontrivial discrepancies between what a simulator can predict and what the student observes in practice. It is only valid for one type of DC motors, the motor in field control mode. It does not support DC motors in armature control modes commonly used in building laboratory servo-trainers.

Associating laboratory experience with theoretical lectures is an important and challenging task especially at early stages of control education. It facilitates a student's transition to more senior and graduate level control concepts. It also provides an effective link to industrial applications. Salient discrepancies between the theoretical and the practical parts at such a fundamental step in control education are detrimental to both goals. These discrepancies will most probably be experienced during the course of a control lab if a first order motor velocity transfer function (2nd order position transfer function) is used as the model of the motor process.

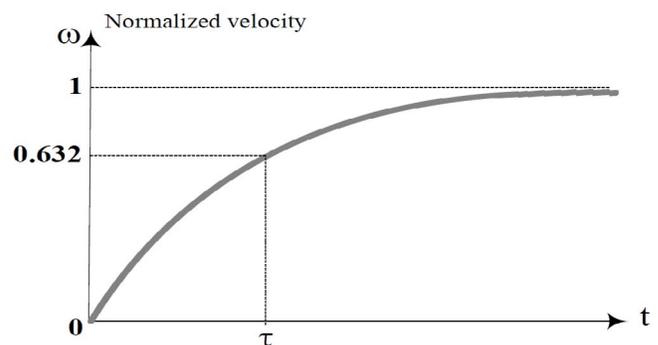

Figure-1: Time constant from a first order step response

Understanding the DC motor's behavior in the laboratory is expected to significantly improve if the second order response is viewed within the confines of a third order one [4]. This makes it important to model a second-order transfer function of the motor's velocity. The procedure commonly taught in 1st control laboratories makes this goal hard to achieve. The students are told to assume that the velocity transfer function is first order with one time constant and a scaling factor equal to the steady state velocity (figure-1).

Advanced techniques exist for obtaining a better model of a motor [5-11]. These techniques are intended as tools of research that do not suit a $1^{st}$ control laboratory. Moreover, to the best of our knowledge, none of these techniques attempt direct experimental identification of the state space model. While there are different methodologies, the common practice seems to first identify the transfer function [13] of the system then construct the state space model using realizations [18].

The only attempt we encountered to introduce direct state space model identification of a generator-motor set in an undergraduate laboratory is that by Basilio and Moreira [12]. The experiment does not use the step response and assumes that the motor's velocity transfer function is first order. The mathematics used is somewhat advanced for a student with basic knowledge in control.

In this paper, we propose two novel, lab-friendly procedures at the undergraduate level for the direct identification of the state space model of a DC motor. The first procedure is based on the pseudo inversion method. It requires a background that is well within what a student in a first control laboratory has. The basic pseudo inversion procedure was originally presented in [19].

In this paper another efficient and novel lab-friendly transfer function identification procedure that is based on the Markov approach is added to the work and the pseudo inversion-based method is presented in more details, especially with regards to experiment implementation, experiment delivery and student assessment.

The second procedure can jointly identify the order of the motor's transfer and estimate its parameters using Markov expansion and state space realizations [18]. However, it requires the student to have a reasonable exposure to relatively advanced tools in the state space approach making it suitable for a second control laboratory. This procedure is novel and was not presented in [19].

Both procedures use only the step response of the motor to construct its state space model. The same procedure and dataset for both methods may be used to compute a second-order or a third-order position transfer function.

While both procedures are thoroughly tested, only the $1^{st}$ procedure, the pseudo inverse-based method, is integrated in the control laboratory syllabus at the EE-department/KFUPM as an experiment. The reason is that the electrical engineering department has only a first control laboratory.

The experiment requires only the following standard equipment, which may be found in any basic control laboratory:

1- Tektronix TDS 2012C , two-channel digital storage oscilloscope (figure-20) + PC interface software
2- Standard laboratory PC with Matlab and MS Excel installed
3- Servo trainer 33-110 from feedback inc. (figure-2)

It supports the following objectives:
1- make experimental state space modeling accessible at the undergraduate level
2- introduce undergraduate students to advanced, control-related mathematical tools
3- strengthening the relation between basic theoretical control concepts and experimental observations
4- introduce a useful, relatively accurate and easy to use experimental modeling tool whose usefulness extends beyond a control laboratory

The experiment was conducted by the students of the EE380 (control systems-I) in the EE department at KFUPM. It was successfully completed in one laboratory session, 2 hours and 45 minutes.

## II. DRAWBACKS of a 2nd ORDER APPROXIMATION

Constant field permanent magnet DC motors are commonly used in building laboratory servo-trainers such as the 33-100. The equivalent circuit and the block diagram of the motor are shown in figure-3. Since a motor of this type can only be controlled through its armature voltage, it is in an armature control mode. A casual justification that is based on the electrical time constant of the motor being much smaller than its mechanical time constant is commonly given to justify the first order velocity transfer function assumption.

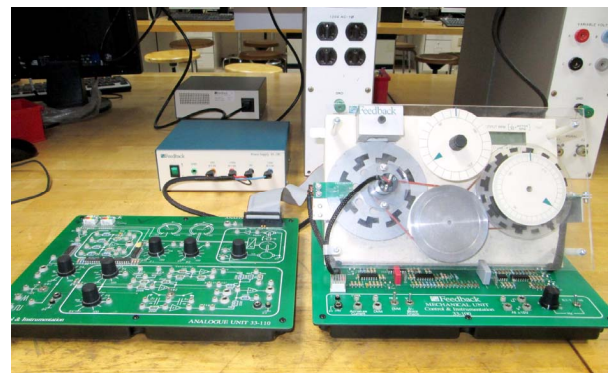

Figure-2: Feedback inc. servo-trainer 33-110

The velocity transfer function (1) of the motor in an armature control mode is:

$$Hv(S) = \frac{K_m}{\tau_e \cdot \tau_m S^2 + (\tau_e + \tau_m)S + (1 + K_b \cdot K_m)} \quad (1)$$

where $\tau_e, \tau_m$ are the electrical and mechanical time constants of the motor respectively, $K_m$ and $K_b$ are constants of the motor. The second order dynamic term has a coefficient that is the multiplication of the electrical and mechanical time constants of the motor. This means that second order dynamics may not be negligible even if the mechanical time constant is considerably bigger than the electrical time constant. Moreover, built-in power electronics circuits such as the servo-amplifier affect the dynamics of the motor.

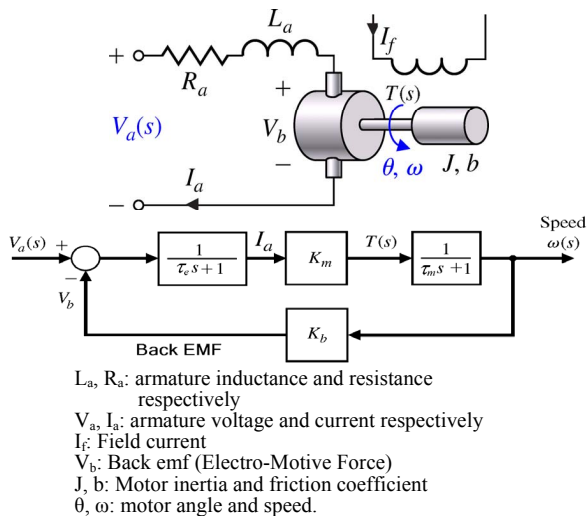

$L_a$, $R_a$: armature inductance and resistance respectively
$V_a$, $I_a$: armature voltage and current respectively
$I_f$: Field current
$V_b$: Back emf (Electro-Motive Force)
J, b: Motor inertia and friction coefficient
θ, ω: motor angle and speed.

Figure-3: DC motor in armature control mode

The usual practice in a 1st, undergraduate control laboratory is to use a second order position transfer function (2) of the first type ($H_p(S)$) to describe a motor's behavior,

$$H_p(S) = \frac{K_m}{S \cdot (S+P)} = \frac{K_m/P}{S \cdot (\tau \cdot S + 1)} \quad (2)$$

where $K_m$ is the coefficient of the motor, -P is its nonzero pole and τ is its time constant. The motor model is then inserted in a standard feedback loop arrangement in order to construct a basic control system (figure-4). The feedback consists of position and velocity feedback loops with nonnegative gains $K_p$ and $K_v$ respectively.

Analysis of the system in figure-4 (theoretical or by simulation) reveals only the three standard modes of behavior: over-damped, critically-damped and under-damped (figure-5).

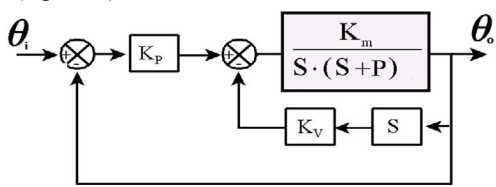

Figure-4: 2nd order servo-motor with position and velocity servo loops.

A second order motor position transfer function model cannot predict system instability. This is due to the fact that second order systems with non-positive poles are unconditionally stable for any negative feedback arrangement. However, in a position control physical experiment, the motor may exhibit sustained oscillations (limit cycles) even when the feedback is negative. In some cases, the cause of oscillations is instability with saturation preventing the magnitude from becoming unbounded (figure-6). This presents the student with a fundamental discrepancy between the theory being taught and the outcome in the laboratory.

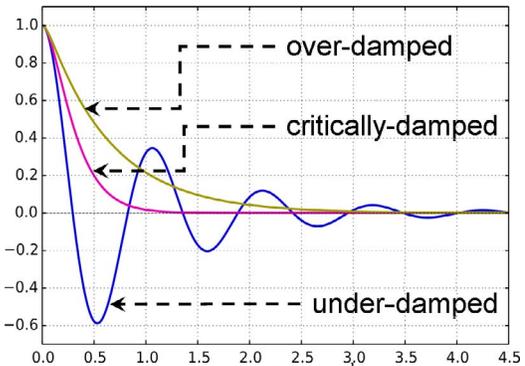

Figure-5: Modes of response of the 2nd order servo-motor in figure-3.

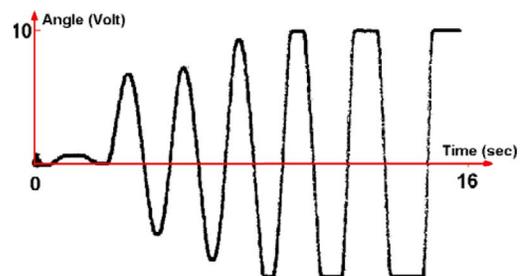

Figure-6: servo-trainer unstable response

At the undergraduate level, the concept of relative stability is tied to the damping coefficient ζ of a second order system (first order velocity transfer function) [2, p.421]. The students are placed under the impression that the servo must exhibit high oscillations (i.e. ζ becomes very low) prior to becoming unstable (figure-7).

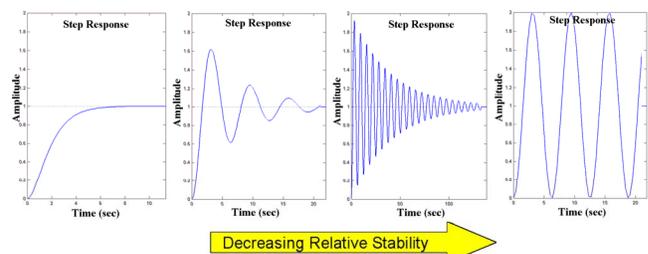

Figure-7: Relative stability based on a second order system.

Third order systems [3] can transit directly from a stable over-damped phase to an unstable phase without exhibiting high oscillations. For example, consider the transfer function in (3) with a parameter α. Changing α from 0.1 to 0.02 (figure-8) converts the response from over-damped to unstable.

$$H(S,\alpha) = \frac{K}{S^3 + \alpha \cdot S^2 + S + .03} \quad (3)$$

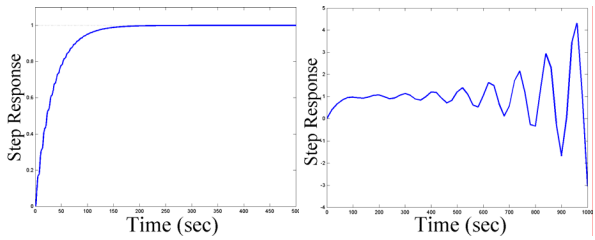

Figure-8: A third order system changing phases

Approximating a 3rd order system by 2nd order one before placing the 3rd order plant in a feedback loop can cause significant errors, even instability, in predicting the response of the motor. Consider a 3rd order open loop system H(S) (4) with a dominant pole that is placed in a closed loop configuration with forward gain K=5 (figure-9).

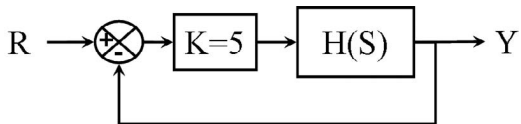

Figure-9: A closed loop position servo.

$$H(S) = \frac{10}{S \cdot (S+1) \cdot (S+10)} \quad (4)$$

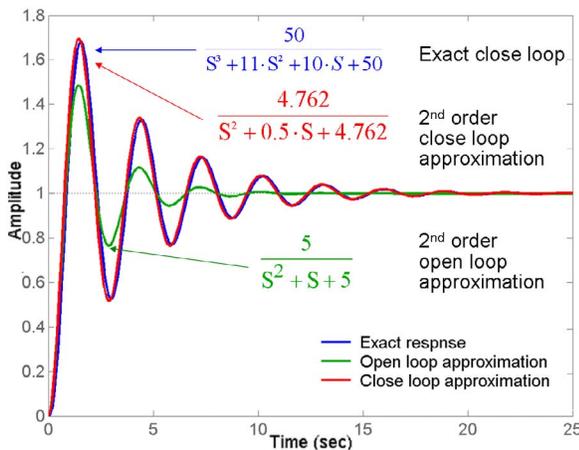

Figure-10: open and closed loop 2$^{nd}$ order approximation versus the exact 3$^{rd}$ order response.

If the quickly fading pole (S=-10) is first removed to produce a 2$^{nd}$ order approximation of H(S) then the approximate is inserted in the feedback loop, the following 2$^{nd}$ order transfer function results for the overall system ($\widetilde{G}(S)$),

$$\widetilde{G}(S) = \frac{5}{S^2 + S + 5} \quad (5)$$

On the other hand, if the exact H(S) is inserted in the feedback loop then the overall system is approximated with a 2$^{nd}$ order system, the overall transfer function is,

$$\widetilde{G}(S) = \frac{4.762}{S^2 + 0.5 \cdot S + 4.762} \quad (6)$$

The step response from the two procedures are compared with the exact one of the third order system (figure-10). As it can be seen, the 2$^{nd}$ order approximation after the insertion of the exact motor model in the feedback loop provides a much more accurate approximation of the exact response.

Experimental testing of motor-based servos in industrial laboratories reveals that at least six modes of operation may be observed when tuning the feedback loop of a physical servo-motor [17]. These modes are summarized in figure-11.

| | Mode | Description | Profile |
|---|---|---|---|
| 1 | Unstable | Instability causes the position to diverge from the reference position in either an oscillatory or an exponential manner. | |
| 2 | Over-damped | A steady and slow motion towards the reference position with no oscillation | |
| 3 | Critically-damped | Fast, steady and oscillation-free motion towards the reference position. | |
| 4 | Under-damped | Oscillations whose strength decays with time are present in motion as it approach the reference position. | |
| 5 | Oscillatory | Sustained position oscillations of equal magnitude where motion does not settle at the reference location. | |
| 6 | Chattering | Audible high-frequency, low-magnitude, sustained oscillations around the reference position are present in motion | |

Figure-11: Mode of operation of an industrial servo-motor.

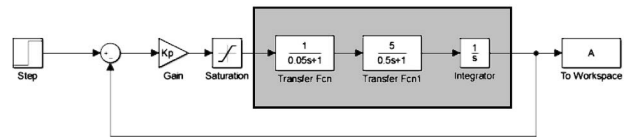

Figure-12: Third order position servo-motor with a saturation nonlinearity.

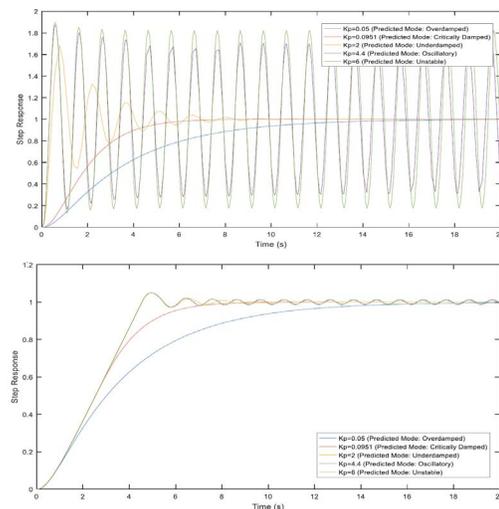

Figure-13: Servo-motor modes predicted by the system in figure-11.

A 3rd order model of the motor coupled with a saturation nonlinearity (figure-12) can reproduce all the modes (figure-13) where a second order model of the motor is not able to predict all the modes.

## III. PSUEDO INVERSION-BASED STATE SPACE IDENTIFICATION

This section presents the mathematical procedure for the computation of the motor's state space model via pseudo inversion. It also presents the supporting operations needed for the procedure to process physical servo data in a reasonably robust manner.

1. The Pseudo inversion approach

In a first control laboratory, it suffices to tell a student that a state space model (7,8) is an alternative to using transfer functions and it has the form:

$$\dot{X} = A \cdot X + B \cdot Va \qquad (7)$$
$$\dot{\theta} = C \cdot X + D \cdot Va \qquad (8)$$

where $X = [\theta \ \dot{\theta} \ Ia]^T$, $C = [0 \ 1 \ 0]$ and $D = [0]$, Ia is the armature current and Va is the motor's armature voltage which is used as the input to control the motor. Since the motor's input is a step function, Va is a constant. The 3x3 matrix A and 3x1 vector B must be determined in order to identify the motor's transfer function. The transfer function is determined using the formula:

$$H(S) = C \cdot (S \cdot I - A)^{-1} B + D \qquad (9)$$

The evaluation may be carried-out using the matlab command: ss2tf(A,B,C,D).

If the signals $\theta(t), \dot{\theta}(t), \ddot{\theta}(t), Ia(t), \dot{Ia}(t)$ can be directly measured or reliably computed from the output of the servo-trainer, equation (7) may be placed in the form (10):

$$\dot{X}(t) = \begin{bmatrix} X^T(t) & 0 & 0 & Va & 0 & 0 \\ 0 & X^T(t) & 0 & 0 & Va & 0 \\ 0 & 0 & X^T(t) & 0 & 0 & Va \end{bmatrix} \cdot \Lambda \qquad (10)$$

or in a more compact manner (11):

$$Z(t) = Q(t) \cdot \Lambda \qquad (11)$$

where $\Lambda = [a_1 \ a_2 \ a_3 \ B^T]^T$ and the a's are the rows of the matrix A

$$A = \begin{bmatrix} a_1 \\ a_2 \\ a_3 \end{bmatrix}. \qquad (12)$$

Sampling (11) at different instants in time $\{t_1,..t_L\}$ to obtain at least L independent measurements (L>12), one may construct the linear system in (13) where the parameters of the motors constitute the unknown vector ($\Lambda$) to be computed:

$$\begin{bmatrix} Z(t_1) \\ \vdots \\ Z(t_L) \end{bmatrix} = \begin{bmatrix} Q(t_1) \\ \vdots \\ Q(t_L) \end{bmatrix} \cdot \Lambda \qquad (13)$$

$\Lambda$ may be computed as:

$$\Lambda = \begin{bmatrix} Q(t_1) \\ \vdots \\ Q(t_L) \end{bmatrix}^+ \cdot \begin{bmatrix} Z(t_1) \\ \vdots \\ Z(t_L) \end{bmatrix} \qquad (14)$$

where the superscript + indicate the Moore-Penrose pseudo inverse. This operation is realized using the matlab command pinv(*). Equation (13) is solved assuming that all coefficients of the A and B matrices are unknown. This may not be necessary since many of the motor's coefficients can be *a priori* determined. When constructing the above system, it ought to be remembered that the initial conditions must be set to zero (i.e. $\theta(0) = 0, \dot{\theta}(0) = 0$ & $Ia(0) = 0.$).

2. Differentiation

The 33-110 servo-trainer provides direct measurements of $\theta(t), \dot{\theta}(t), \& Ia(t)$. Numerical differentiation needs to be applied in order to compute the angular acceleration and the derivative of the armature current. Robust differentiation of natural signals [14] is not easy to teach at the undergraduate level. Also directly using Eulers discritization (15) to compute the derivatives will not produce satisfactory results

$$\ddot{\theta}(t) \approx \frac{\dot{\theta}(t) - \dot{\theta}(t - \Delta T)}{\Delta T}. \qquad (15)$$

There are reasonably accurate differentiation formulae that are usable by an undergraduate student. For example, the formula in (16) produces good results. Better accuracy may be obtained from formula (17) [20, Table 25.2].

$$\ddot{\theta}(t) \approx \frac{\dot{\theta}(t + \Delta T) - \dot{\theta}(t - \Delta T)}{2 \cdot \Delta T} \qquad (16)$$

$$\ddot{\theta}(t) \approx \frac{-\dot{\theta}(t + 2 \cdot \Delta T) + 8 \cdot \dot{\theta}(t + \Delta T) - 8 \cdot \dot{\theta}(t - \Delta T) + \dot{\theta}(t - 2 \cdot \Delta T)}{12 \cdot \Delta T} \qquad (17)$$

3. Noise reduction

Natural signals such as those produced by the servo-trainers are usually noisy. For the DC motor, the source of noise is the power electronics and the signal encoders on-board the trainer. While many effective techniques exist for noise removal [5], they may not be suitable for use by students in a 1st control laboratory. Moreover, these techniques assume an additive white Gaussian noise (AWGN) which is most probably not the case with the servo trainer. Performing noise removal is subject to stringent requirements at such a basic level. Ease of application and use by the students is a fundamental requirement. A crucial property the noise filtering procedure must have is that it must not distort the registration of the system dynamics in the measured signals. This rules-out the use of simple IIR and FIR filters which exist as a single Matlab command.

Here, a sample reduction approach is suggested to reduce the effect of noise without disturbing the informational content of the servo signals. Sampling a signal every T seconds is equivalent to applying a Nyquist lowpass filter [21] in the frequency domain with a bandwidth W=1/T (figure-14).

Sample reduction reduces W and increases T. This causes the removal of high frequency components without affecting the informational content of the measured signal (figure-15).

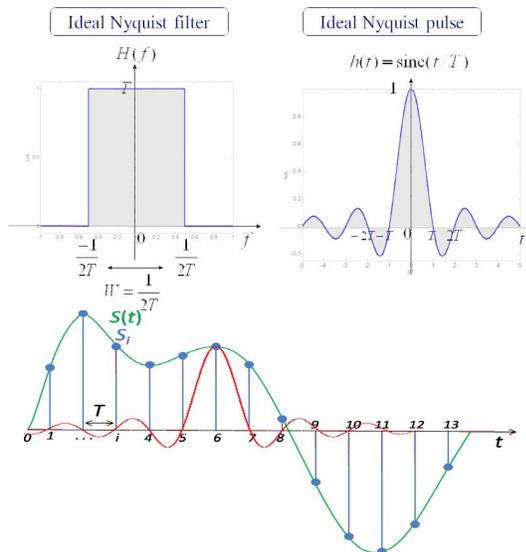

Figure-14: Nyquist filtering inherent in signal sampling.

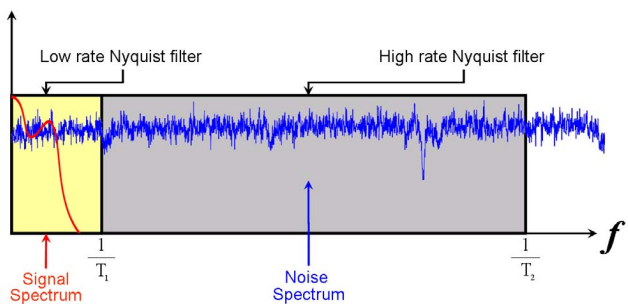

Figure-15: Noise filtering through sample reduction.

A 4 ms data record contains close to 2000 samples. Only 12 samples are needed to uniquely solve equation (13) for a third order model. Therefore, a significant reduction in bandwidth of about 100 times can be achieved. Figure-16 shows the speed signal record with 1500 samples and sampling period of 4 ms.

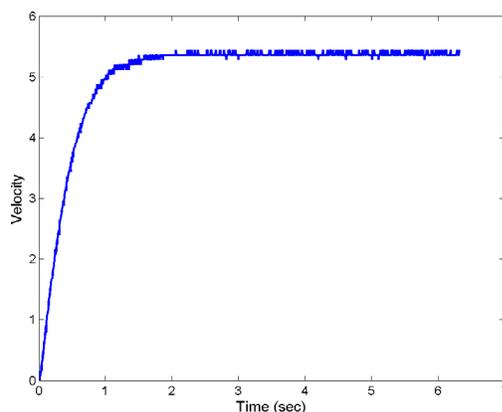

Figure-16: velocity step response of the motor

The acceleration is shown in figure-17 for the two cases of direct differentiation using (17) and differentiation after 1:30 under sampling ratio. As can be seen, the effect of noise on the acceleration signal greatly diminished.

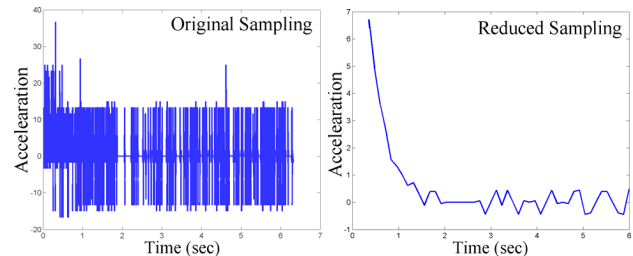

Figure-17: reducing noise effect on differentiation by sample rate reduction

An H-bridge that uses pulse-width modulation seems to be used by the servo trainer to control velocity. This causes the armature current to be highly noisy (figure-18). While advanced signal processing techniques can de-noise the current, the sample reduction method is still able to produce acceptable results.

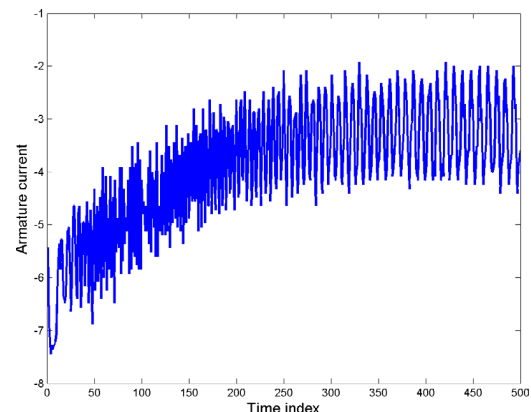

Figure-18: High noise level in armature current

## IV. MARKOV-BASED STATE SPACE IDENTIFICATION

The Markov approach is an effective means for constructing minimal state space realizations from transfer functions [18]. To the best of our knowledge, the technique was not used before for the direct construction of a system state space model from experimental data. This section describes a mathematical procedure to utilize this approach for the determination of the motor's state space model from its velocity step input.

The step position response of the motor in the S-domain may be written as (18):

$$X(S) = H(S) \cdot \frac{V_0}{S}. \quad (18)$$

The velocity step response in the S-domain is (19):

$$V(S) = S \cdot X(S) = S \cdot H(S) \cdot \frac{V_0}{S} = H(S) \cdot V_0 \quad (19)$$

As can be seen, the velocity step response of the motor in the Laplace domain is equal to its transfer function scaled by the magnitude of the step input. Since the transfer function of a DC motor is strictly proper it may be written using the Markov parameters (q(i)) as (20):

$$H(S) = \sum_{i=1}^{\infty} q(i) \cdot S^{-i} \quad (20)$$

Therefore, the velocity step response of the motor in the time domain may be written as (21):

$$v(t) = V_0 \cdot l^{-1}(H(S)) = V_0 \cdot \sum_{i=1}^{\infty} q(i) \cdot \frac{t^{i-1}}{(i-1)!} \quad (21)$$

Computing an infinite number of Markov parameters in order to indentify the system is neither possible nor necessary for that matter. A finite number of parameters is sufficient to determine both the system order and construct a minimal state space model of the motor. Consider an approximation of the transfer function that is constructed from a finite number (Lm) of Markov parameters (22):

$$\hat{H}(S) = \sum_{i=1}^{Lm} q(i) \cdot S^{-i} \quad (22)$$

The approximate velocity step response is (23):

$$\hat{v}(t) = V_0 \cdot \sum_{i=1}^{Lm} q(i) \cdot \frac{t^{i-1}}{(i-1)!} \quad (23)$$

Notice that

$$\lim_{S \to \infty} H(S) = \lim_{S \to \infty} \hat{H}(S) \quad (24)$$

This implies that

$$\lim_{t \to 0} v(t) = \lim_{t \to 0} \hat{v}(t) \quad (25)$$

Since the approximate velocity step response is both continuous and rapidly converging in terms of the Markov parameter number, one may find a short time interval Tε such that (26):

$$|v(t) - \hat{v}(t)| \leq \delta \quad \forall t \leq T\varepsilon \quad (26)$$

where δ is arbitrarily small. For example, consider the system with transfer function (27)

$$H(S) = \frac{36}{S^2 + S + 36} \quad (27)$$

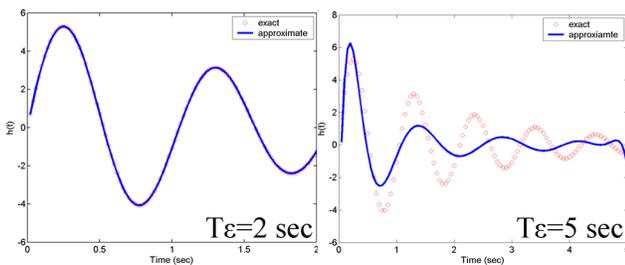

Figure-19: Effect of Tε on the accuracy of Markov-based response construction.

H(S) is approximated using 11 Markov parameters. Two values of the Tε are used to construct the impulse response, 2 seconds and 5 seconds. The impulse response for the two cases is plotted along with the exact one (figure-19). As can be seen for Tε=2 sec the approximate impulse response is almost identical to the exact one. For Tε=5 major deviations appeared between the approximate and the exact for the whole time axis and were not restricted for t>2 sec.

To compute the Lm Markov parameters, N samples of v(t) are obtained in the period Tε and the following linear system is constructed

$$\begin{bmatrix} v(1 \cdot \Delta) \\ v(2 \cdot \Delta) \\ \vdots \\ v(N \cdot \Delta) \end{bmatrix} = \begin{bmatrix} 1 & \Delta & .. & \frac{\Delta^{Lm-1}}{(Lm-1)!} \\ 1 & 2\Delta & .. & \frac{(2\Delta)^{Lm-1}}{(Lm-1)!} \\ \vdots & \vdots & .. & \vdots \\ 1 & N\Delta & .. & \frac{(N\Delta)^{Lm-1}}{(Lm-1)!} \end{bmatrix} \begin{bmatrix} q(1) \\ q(2) \\ \vdots \\ q(L) \end{bmatrix} \quad (28)$$

The pseudo inverse of the above matrix may be used to compute the Lm Markov parameter vector. Notice that the matrix is full column rank. However, in order for the system to have a solution, the initial conditions must be zero (i.e. v(0)=0).

Once the Markov parameters are computed, the Hankel matrix (Ha) is constructed (29). The number of Markov parameters needed to construct a KxK Hankel matrix is Lm=2K-1.

$$Ha(K,K) \begin{bmatrix} q(1) & q(2) & .. & q(K) \\ q(2) & q(3) & .. & q(K+1) \\ \vdots & \vdots & .. & \vdots \\ q(K) & q(K+1) & .. & q(2K-1) \end{bmatrix} \quad (29)$$

Ideally the rank of the Hankel matrix (M) is equal to the order of the system. Despite the approximation, the rank of the Hankel matrix can still give a good idea about the order of the system. For example, performing singular value decomposition on the 6x6 Hankel matrix constructed from the 11 markov parameters of transfer function, yields the following normalized Eigen values: 1, 0.29, .016, 0,0,0. As it can be seen the Eigen values after the second are almost negligible. This is a strong indicator that the system is second order. Performing the same thing for the first order transfer function (30)

$$H(S) = \frac{1}{S+1} \quad (30)$$

yields the normalized Eigen values: 1, .088, .0043,0,0,0. The Eigen values became negligible after the first one strongly indicating a first order system.

$$\tilde{H}a(M,M) \begin{bmatrix} q(2) & q(3) & .. & q(M+1) \\ q(3) & q(4) & .. & q(M+2) \\ \vdots & \vdots & .. & \vdots \\ q(M+1) & q(M+2) & .. & q(2M) \end{bmatrix} \quad (31)$$

The dimension of the Hankel matrix K should be selected so that K>M or equivalently Lm>2M-1. A Reduced in size

Hankel matrix, Ha(M,M) has to be constructed along with a shifted Hankel matrix of size MxM ($\tilde{\mathrm{Ha}}$(M,M)),

The minimal dimension Hankel matrix may be expressed as:
$$\mathrm{Ha}(M,M) = O \cdot C \quad (32)$$
Where $O$ is the observability matrix of the state space system and $C$ is its controllability matrix. The shifted Hanke matrix of reduced dimensions may be written as
$$\tilde{\mathrm{Ha}}(M,M) = O \cdot A \cdot C \quad (33)$$
where A is the state matrix. Equations 32 and 33 are the basis for generating [18] a large number of minimal state space realizations (e.g. companion forms, balanced realizations etc.). Since all minimal realizations of a system are similar, they all produce the same transfer functions. In addition, there exist a full rank matrix (similarity transform) that can convert each realization to another.

A minimal state space realization in companion form is easily constructed by choosing the observability matrix to be the identity matrix (I)
$$O = I \quad (34)$$
As a result, the state matrix A may be computed as (35)
$$A = \tilde{\mathrm{Ha}}(M,M) \cdot \mathrm{Ha}(M,M)^{-1} \quad (35)$$
The B matrix is computed as the first column of Ha(M,M).
$$B = \begin{bmatrix} g(1) & g(2) & .. & g(M) \end{bmatrix}^T \quad (36)$$
The C matrix is
$$C = \begin{bmatrix} 1 & 0 & .. & 0 \end{bmatrix} \quad (37)$$
and the D matrix is zero (D=[0]).

The motor's transfer function can be uniquely generated from the state space model.

Both procedures described above are lab-friendly and doable using standard control laboratory equipment. However, the Markov approach requires relatively advanced knowledge of state space that is usually taught at the senior undergraduate level or in a 1st graduate course in control. Unfortunately, the control laboratory curriculum of the EE department at KFUPM has only a 1st control laboratory. We strongly believe that the pseudo inversion-based procedure does not only suit a 1st control laboratory, but also do serve critical objectives that are beneficial to the students at such an early stage in their control education.

## V. DIRECT INVERSION: EXPERIMENT PROCEDURE & DATA CONDITIONING

This section describes the different supporting procedures needed to implement the state space identification method using the provided laboratory equipment along with the workflow the student has to follow in order to perform the experiment.

1. Data synchronization:
A basic two-input digital oscilloscope such as the Tektronix TDS 2012c (figure-20) is selected for recording the data instead of a multi-input oscilloscope or data acquisition card. This type of oscilloscopes is affordable and may be found in any undergraduate control laboratory. The oscilloscope can only simultaneously acquire two data channels at a measurement.

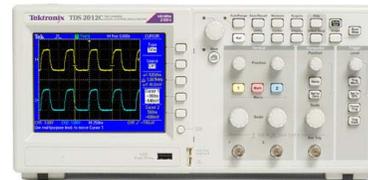

Figure-20: The TDS 2012c digital oscilloscope used in the experiment.

The TDS 2012c stores the sensed data in an Excel sheet (figure-21). This makes it possible to produce synchronized multi-data streams one signal at a time. Each measurement is performed separately and stored in an Excel sheet that contains the time trace, step input and the quantity that is measured. The data from each sheet is truncated at the instant the step input changes value from zero to full voltage. This is considered to be the common zero time for all records. Excess data should also be removed so that all records are the same length. All the signals can then be copied to a common Excel sheet that contains the synchronized servo-signals of the motor.

Figure-21: Excel sheet from digital oscilloscope.

2. Position measurements unwrapping:
The position encoder on the servo trainer 33-110 experiences a sudden jump from -10 volts to +10 volts and vice versa if it rotates more than 360 degrees (figure-22). The situation requires special care due to the presence of low voltage signals (mostly zeros) in-between jumps. A simple logical condition that is provided to the students may be used to reliably detect those jumps. Ideally, if a sudden change in encoder voltage from positive to negative is detected and the value of the position signal is close to 10 volts, the difference in value is added to the subsequent samples.

After the position signal is unwrapped, the whole data record is shifted so the first sample value is zero.

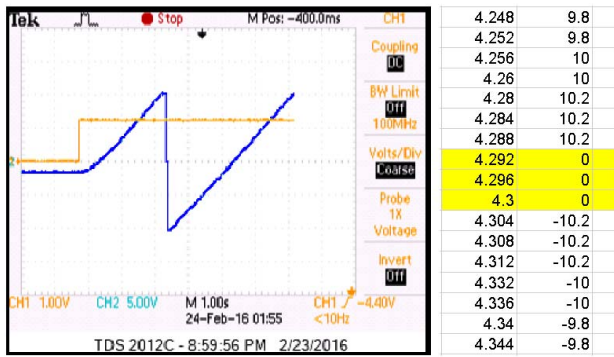

Figure-22: sudden jumps in position caused by the encoder

3. Data smoothing & Differentiation:
Data acquisition should continue until the speed of the motor settles at a steady state value. Signal sampling is carried out at a relatively high rate of 2 or 4 ms per sample. The data record must be under-sampled in order to reduce the signal-sampling rate before differentiating it. A sampling period of 100 ms can produce good smoothing effect on the data and permit reliable computation of the state space model parameters. The reduced sample record is differentiated using equation (17). The first and the last two samples of the record are discarded.

4. From state space to Transfer function:
Equation (38) shows a typical result of the identification method for a position and velocity only state vector. For a 1st order velocity transfer function, the first row of the A matrix should be [0 1]. The first element of the B matrix should be [0]. Numerical accuracy causes the computed values to experience some deviation from the ideal ones. The matlab command ss2tf(A,B,C,D) produces the velocity transfer function shown in (39). Selecting C=[1 0] produces the position transfer function in (40).

$$A = \begin{bmatrix} .0042 & 1.0325 \\ -.0327 & -2.3145 \end{bmatrix}, \quad B = \begin{bmatrix} -.0371 \\ 2.1751 \end{bmatrix} \quad (38)$$
$$C = \begin{bmatrix} 0 & 1 \end{bmatrix}, \quad D = [0]$$

$$Hv(S) = \frac{2.1751 \cdot S + .0103}{S^2 + 2.3187 \cdot S + .0435} = \frac{2.1751 \cdot (S + .0048)}{(S + .0189) \cdot (S + 2.2998)} \quad (39)$$

$$Hp(S) = \frac{2.1751 \cdot S + 0.0104}{S^2 + 2.3187 \cdot S + .0435} \quad (40)$$

As can be seen, the velocity transfer function has a pole and a zero that are very close to the origin. They are caused by numerical issues of the identification method. They may be cancelled to obtain the approximate velocity and position transfer functions in equations (41) and (42) respectively

$$\tilde{H}v(S) = \frac{2.1751}{S + 2.2998} \quad (41)$$

$$\tilde{H}p(S) = \frac{2.1751}{S \cdot (S + 2.2998)} \quad (42)$$

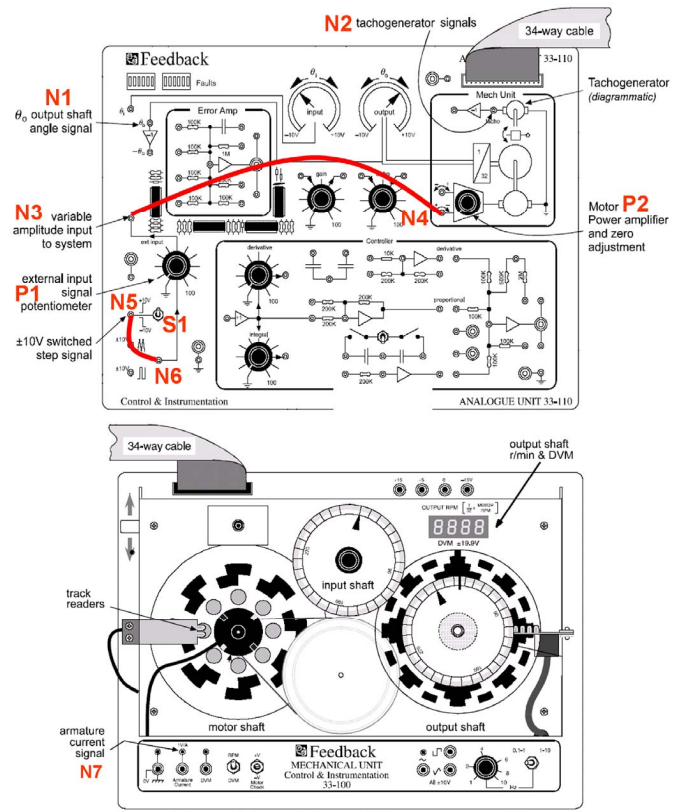

Figure-23: Connection diagram

5. Experimental workflow:
Setp-01: While the servo-trainer board is unconnected (figure-23), adjust the potentiometer of the power (P2) so that the motor is not moving,
Step-02: Connect N5 to N6 in order to select the step input option from the voltage source on-bard the trainer. Make sure that switch-1 (S1) is at the middle (zero) position.
Step-03: Connect the variable output of the step source (N3) to the servo-amplifier of the motor (N4). The value of the input is controlled by the potentiometer P1.
Step-04: Set the oscilloscope to a suitable time-base and do not change it. The time-base controls the sampling rate at which data is stored in the oscilloscope.
Step-05: Connect channel-1 of the digital oscilloscope to N3 and channel-2 to the angle of the motor (N1).
Step-06: Apply a step input by switching S1 to the 10 V side directly after pushing the record button on the oscilloscope. 1.5 seconds or more of data are needed for the speed to settle
Step-07: Store the oscilloscope collected data in an EXCEL sheet (automatically performed by the software of the oscilloscope)
Step-08: Keep the magnitude of the step input (i.e. the position of P1) unchanged.

Step-09: Measure the velocity step input of the motor by Connecting chanel-2 of the oscilloscope to the tachogenerator output (N2). Repeat steps 5 & 6

Step-10 Record the final speed of the motor. The final speed is displayed digitally on the mechanical unit

Step-11: Don't change the magnitude of the step input (i.e. the position of P1)

Step-12: Connect chanel-2 of the oscilloscope to the armature current output (N7). Repeat steps 5 & 6.

6. Experiment Delivery

A student who took the 1st control system course at the EE department a semester before the development of the experiment started was enlisted. His presence was helpful in integrating the students' perspective as a factor that shaped the format of the experiment.

The experiment is about six pages and contains the traditional components of objectives, equipment, introduction, experimental procedure, analysis and guidelines to write the report. The introduction provides the background theory needed to understand and perform the experiment. While the basics of the theory are discussed, the related Matlab procedure is described in clear details. To perform the needed functions, e.g. phase unwrapping of the position signal, snippets of the needed Matlab code are given to the students (figure-24).

Since the experiment strongly depends on being familiar with the Tektronix TDS 2012C oscilloscope, step by step handouts describing how to use the needed oscilloscope functions were generated and distributed in advance to the students.

```
for i=2:L;
   if(abs(P(i))< δ  && abs(P(i)-P(i-1))>5); P(i)=P(i-1); end;
end
```

```
for i=1:L-1
   if((P(i)-P(i+1))>+15); P(i+1:L)=P(i+1:L)+20.4; end
   if((P(i)-P(i+1))<-15); P(i+1:L)=P(i+1:L)-20.4; end
end
```

Figure-24: Matlab code snippets provided to the students for unwrapping the position encoder signal

Since the experiment is designed assuming only minor exposure to the state space approach, the state space vector is restricted to the velocity and position only. The students were encouraged to experiment with the complete state vector with the armature current added. They were also made aware that the same approach can be used in the circuit lab to determine the state space model of RC/RLC circuits

## VI. RESULTS

In this section, the two state space identification procedures are thoroughly tested. The stability and repeatability of the pseudo inverse-based identification method is extensively examined under laboratory conditions using physical signals obtained from the servo-trainer. The Markov-based procedure is explored using both synthetic data with known base-truth and physical signals from the servo-trainer.

1- Pseudo Inversion:

The results in this section demonstrate the stability of the pseudo inversion procedure and its suitability to be performed by undergraduate students in a 1st control laboratory. Emphasis is placed on the reduced state space model with position and velocity as the state vector. The experiment's ability to identity the extended model of the motor with position, velocity and armature current is also demonstrated.

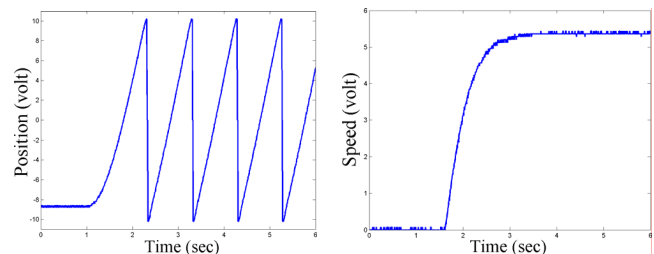

Figure-25: raw position and velocity step response of the motor

A. A typical case:

Selecting the step signal to have a magnitude of 5.92 volts permits the servo-trainer motor to operate in the linear region and avoids significant artifacts in the recorded position and velocity signals which are shown in figure-25. The output of the tacho-generator settles at 5.36 volt which corresponds to 1824 revolution per minute (rpm). A sampling period of 4 ms is used to collect 6 seconds of data. The speed response is used to directly estimate a 2.1354 coefficient of the motor and a 0.4241 seconds time constant. The resulting first order velocity transfer function (43) is:

$$Hv(S) = \frac{2.1354}{S + 2.3579} \quad (43)$$

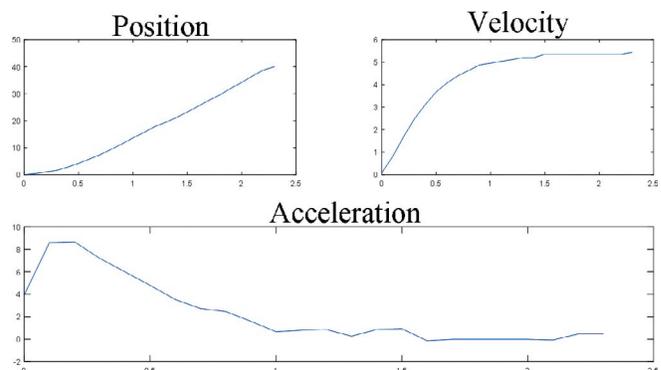

Figure-26: Under-sampled position, velocity and acceleration signals

Figure-26 shows a sample-reduced data record of 2.4 seconds of the unwrapped position, speed and acceleration obtained by differentiating the speed using formula (17).

The sampling rate is reduced 25 times yielding a sampling period of 100 ms. The state space system that results from solving equation (13) is shown in equation (44) and equation (45) shows the corresponding velocity transfer function. The state space generated transfer function yields a time constant of 0.4428 seconds and a motor coefficient of 2.1685. Figure-27 shows the experimental velocity step response along with the one generated by the estimated transfer function. As can be seen, the two responses are reasonably close to each other.

$$A = \begin{bmatrix} 0.0 & 1.00 \\ -.0192 & -2.267 \end{bmatrix}, \quad B = \begin{bmatrix} 0.00 \\ 2.1685 \end{bmatrix} \quad (44)$$
$$C = \begin{bmatrix} 0 & 1 \end{bmatrix}$$

$$Hv(S) = \frac{2.1685 \cdot S}{S^2 + 2.2585 \cdot S + .0192} \approx \frac{2.1685}{S + 2.2585} \quad (45)$$

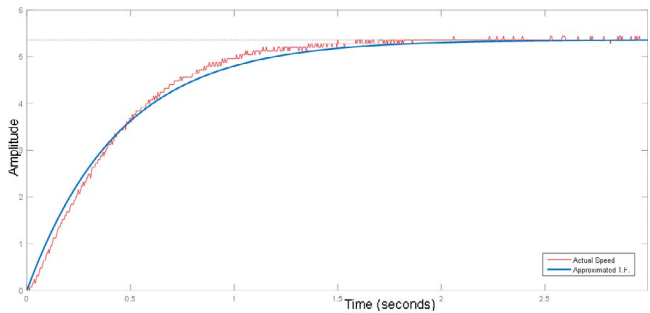

Figure-27: Velocity step response from TF and actual velocity response

Table-1 shows the results for a step input with magnitude 5.92 volts. The table shows the effect of the record length and sample reduction ratio on the motor's estimated time constant and coefficient. As can be seen, for a sample reduction ratio above 60 ms (1:15), the effect is negligible for both sample ratio reduction and record length.

B. Sensitivity to reference input

The nonlinearities in the servo-trainer (e.g. deadzone, static and colomb frictions, saturations, etc.) have the potential to affect the stability of the experiment. Here this issue is tested under a wide range of critical experiment parameters. The magnitude of the step input is varied from low value where the effect of dead zone and static friction nonlinearities is most pronounced to a high value where saturation can significantly impact system behavior. The results for the input Vin=5.92 is repeated for the low input of Vin=1.24 V and the high one of Vin=8.8 V. The results remained reasonably stable and are shown in tables-2 and table-3 respectively for both inputs.

C. Extended state space model

Sample reduction can reasonably suppress the H-bridge induced armature current noise [16]. It is, never the less, observed that how well the model fits the experimental results is dependent on the magnitude of the step input. A signal duration of 4.3 seconds at a sampling period of 4 ms and an under-sampling ratio of 1:30 are used to estimate the extended ($X = \begin{bmatrix} \theta & \dot\theta & Ia \end{bmatrix}^T$.) state space model. The magnitude of the step input is taken to be Va=5.92 volt and the result is shown in (46). The experimental velocity response and the one produced by the estimated velocity transfer function are shown in figure-28. A good match between the two is observed despite the significant noise level in the armature current.

Table-1: Effect of record length and sample reduction ration on the estimated transfer function, Vin=5.92

| ΔT | Record Length | | |
|---|---|---|---|
| | 1.9 s | 2.4 s | 2.9 s |
| 40 ms | k = 1.8045<br>τ = 0.5018 | k = 1.888<br>τ = 0.4795 | k = 1.9407<br>τ = 0.4665 |
| 60 ms | k = 1.8821<br>τ = 0.4811 | k = 1.955<br>τ = 0.4631 | k = 1.9677<br>τ = 0.4601 |
| 80 ms | k = 1.8334<br>τ = 0.4939 | k = 1.9378<br>τ = 0.4672 | k = 1.996<br>τ = 0.4536 |
| 100 ms | k = 1.9816<br>τ = 0.4569 | k = 2.1685<br>τ = 0.4428 | k = 2.0967<br>τ = 0.4318 |
| 120 ms | k = 2.1698<br>τ = 0.4173 | k = 2.2031<br>τ = 0.4110 | k = 2.2338<br>τ = 0.4053 |

Table-2: Effect of record length and sample reduction ration on the estimated transfer function, Vin=1.24

| ΔT | Record Length | | |
|---|---|---|---|
| | 1.9 s | 2.4 s | 2.9 s |
| 40 ms | k = 2.2495<br>τ = 0.4535 | k = 2.4533<br>τ = 0.4471 | k = 2.3842<br>τ = 0.4600 |
| 60 ms | k = 2.0097<br>τ = 0.5457 | k = 2.2977<br>τ = 0.4773 | k = 2.2408<br>τ = 0.4894 |
| 80 ms | k = 1.6803<br>τ = 0.6527 | k = 2.0643<br>τ = 0.5313 | k = 2.0340<br>τ = 0.5392 |
| 100 ms | k = 1.9750<br>τ = 0.5553 | k = 2.3073<br>τ = 0.4753 | k = 2.2075<br>τ = 0.4968 |
| 120 ms | k = 2.5971<br>τ = 0.4223 | k = 2.6418<br>τ = 0.4152 | k = 2.6724<br>τ = 0.4104 |

Table-3: Effect of record length and sample reduction ration on the estimated transfer function, Vin=8.8

| ΔT | Record Length | | |
|---|---|---|---|
| | 1.9 s | 2.4 s | 2.9 s |
| 40 ms | k = 1.7977<br>τ = 0.4804 | k = 1.7722<br>τ = 0.4873 | k = 1.9158<br>τ = 0.4508 |
| 60 ms | k = 1.6849<br>τ = 0.5126 | k = 1.7839<br>τ = 0.4841 | k = 1.8632<br>τ = 0.4635 |
| 80 ms | k = 1.6324<br>τ = 0.5291 | k = 1.8767<br>τ = 0.4602 | k = 1.8456<br>τ = 0.4679 |
| 100 ms | k = 1.8662<br>τ = 0.4628 | k = 1.9350<br>τ = 0.4463 | k = 1.9871<br>τ = 0.4346 |
| 120 ms | k = 1.9026<br>τ = 0.4539 | k = 2.1308<br>τ = 0.4053 | k = 2.0611<br>τ = 0.4190 |

$$A = \begin{bmatrix} 0 & 1 & 0 \\ -.0101 & -2.4731 & .6196 \\ -.0031 & -.0717 & -.2191 \end{bmatrix}, \quad B = \begin{bmatrix} 0 \\ 2.3373 \\ .082 \end{bmatrix} \quad (46)$$

$$C = \{0 \quad 1 \quad 0\}$$

$$H_v(S) \approx \frac{2.337 \cdot S + .5630}{S^2 + 2.692 \cdot S + .597}$$

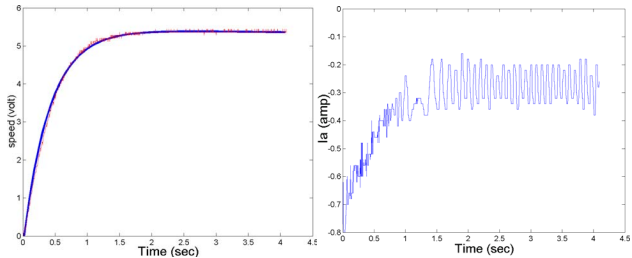

Figure-28: measured velocity step response, the fitted response and the corresponding armature current, Va=5.92.

Equation (47) shows the experimental state space model and approximate velocity transfer function for the low step input Va: 1.24. A sample reduction ratio of 1:43 is used. Figure-29 shows the measured velocity step response, the fitted response from the approximate transfer function along side the armature current. Despite the high noise in the current the fit is reasonably accurate and the model is close to the case of Va=5.92.

$$A = \begin{bmatrix} 0 & 1 & 0 \\ -.0140 & -2.0251 & -.2888 \\ -.0079 & -.0601 & -.2368 \end{bmatrix}, \quad B = \begin{bmatrix} 0 \\ 2.3419 \\ -.0459 \end{bmatrix} \quad (47)$$

$$C = \{0 \quad 1 \quad 0\}$$

$$H_v(S) \approx \frac{2.318 \cdot S + .534}{S^2 + 2.2619 \cdot S + .476}$$

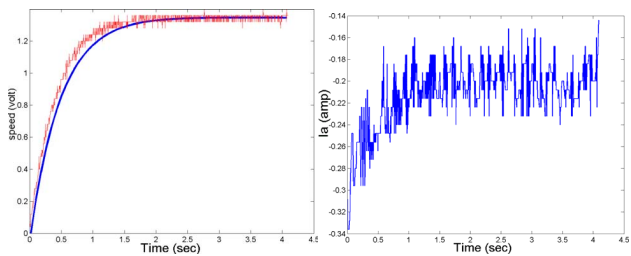

Figure-29: measured velocity step response, the fitted response and the corresponding armature current, Va=1.24.

The results for a high step input (Va=8.8) are shown in equation (48) and figure-30. A sample reduction ratio of 1:18 is used. The results are still reasonably accurate.

$$A = \begin{bmatrix} 0 & 1 & 0 \\ -.0055 & -2.1023 & -3.795 \\ -.0039 & -.1000 & -.4132 \end{bmatrix}, \quad B = \begin{bmatrix} 0 \\ 1.7223 \\ .0979 \end{bmatrix} \quad (48)$$

$$C = \{0 \quad 1 \quad 0\}$$

$$H_v(S) \approx \frac{1.7223 \cdot S + .3402}{S^2 + 2.5156 \cdot S + .4946} \quad (49)$$

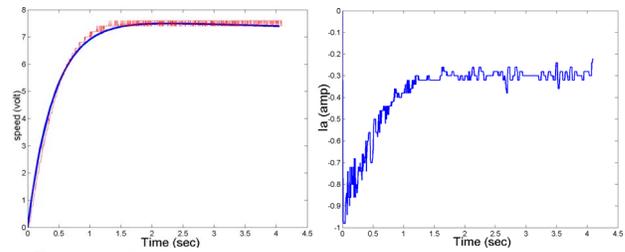

Figure-30: measured velocity step response, the fitted response and the corresponding armature current, Va=8.8.

Currently a low complexity, FFT-based procedure is being developed for removing the noise components if the armature current without disturbing the information-bearing elements of the signal (figure-31). The procedure has a promising ability to suppress the armature current noise (figure-32). It is tested for the Va=5.92 case using the same 1:30 sample reduction ratio. The resulting state space model and approximate transfer function are shown in equation (50). As it can be seen the obtained velocity transfer function is close to the one in (46) and the response of the transfer function fits well the experimental data (figure-33). However, more work is still needed before it becomes suitable for use in a first laboratory in control.

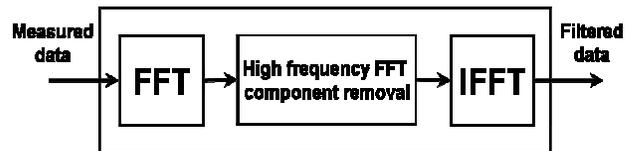

Figure-31: FFT-based denoising filter

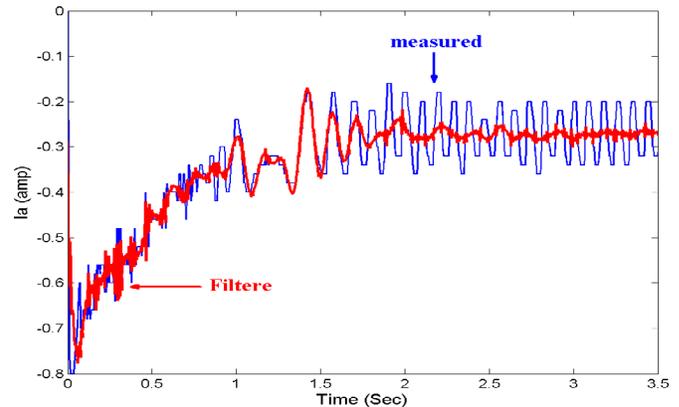

Figure-32: FFT-filtered armature current along with the raw measurement

$$A = \begin{bmatrix} 0 & 1 & 0 \\ -.006 & -2.314 & -.2987 \\ -.0007 & -.1504 & -.2632 \end{bmatrix}, \quad B = \begin{bmatrix} 0 \\ 2.1322 \\ .135 \end{bmatrix} \quad (50)$$

$$C = \{0 \quad 1 \quad 0\}$$

$$H_v(S) \approx \frac{2.1322 \cdot S + .5209}{S^2 + 2.5772 \cdot S + .5703} \quad (51)$$

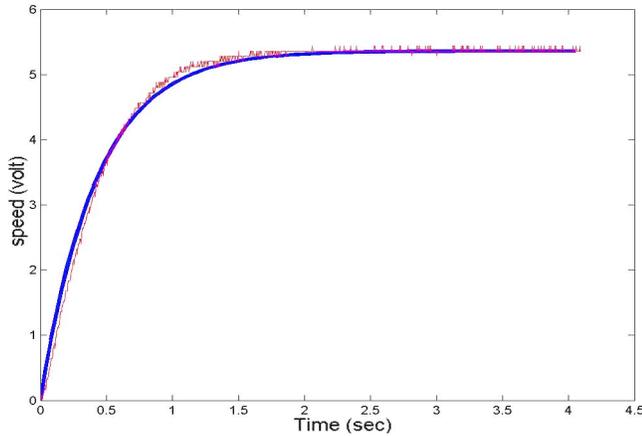

Figure-33: velocity step response from transfer function and experimental data for the extended state space model

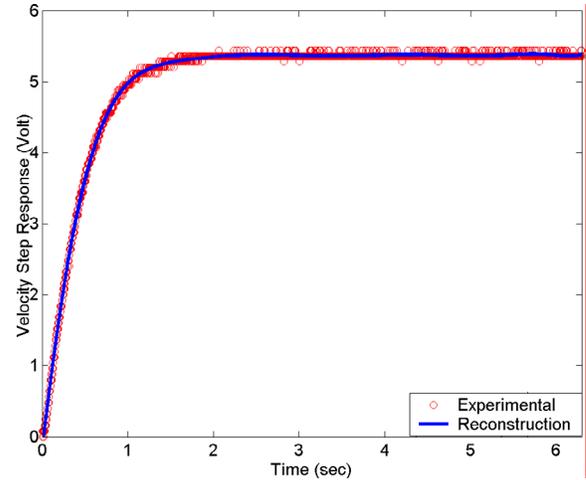

Figure-34: Experimental speed step response and the one reconstructed from the Markov parameters.

2. Markov-based Identification
This section demonstrates the feasibility of the Markov approach as an experimental tool for jointly identifying the transfer function of a motor along with determining its order in the laboratory. The approach is applied to synthetic data where the transfer function is *a priori* known as well as the experimental data obtained from the servo-trainer.

A. Synthetic example
This example demonstrates the ability of the Markov approach to determine jointly the order and coefficients of a motor's transfer function using its speed response only. The synthetic transfer function in (27) is used as the base-truth. The impulse response of the transfer function is used to estimate 11 Markov parameters. The interval on which the estimation is done is: $T\varepsilon=0.1$ sec and a sampling rate of $T\varepsilon/50=.002$ sec (2 ms) is used. The parameters are used to construct a 6x6 Hankel matrix. The rank of this matrix is two. Therefore, it is determined that only 2x2 Hankel and shifted Hankel matrices (52) are needed to fully identity the transfer function.

$$\text{Ha} = \begin{bmatrix} 0 & 36 \\ 36 & -36.001 \end{bmatrix}, \quad \widetilde{\text{Ha}} = \begin{Bmatrix} 36 & -36.001 \\ -36.001 & -1260 \end{Bmatrix} \quad (52)$$

The resulting state matrix A is

$$A = \widetilde{\text{Ha}} \cdot \text{Ha}^{-1} = \begin{bmatrix} 0 & 1 \\ -35.995 & -1 \end{bmatrix}. \quad (53)$$

The B and C matrices are:

$$B = \begin{bmatrix} 0 & 36 \end{bmatrix}^T, \quad C = \begin{bmatrix} 1 & 0 \end{bmatrix}. \quad (54)$$

The estimated transfer function is

$$H(S) = \frac{36}{S^2 + S + 35.995} \quad (55)$$

As can be seen the Markov approach accurately estimated the transfer function. However, one must keep in mind that the noise-free synthetic nature of the signal is a major contributor to this accuracy. Experimental signals contain noise that does limit estimation accuracy.

B. Experimental motor example
In this subsection, the velocity response obtained from the servo-trainer at the input level 5.92 volts is processed using the Markov approach in order to identify the transfer function of the motor. Eleven Markov parameters provided tight approximation of the whole experimental velocity step response data record, $T\varepsilon=6.3$ seconds, (figure-34).

A singular value decomposition of the 6x6 Hankel matrix produced the following normalized Eigen values: 1.0, 0.426, 0.1, 0.009, 0.0009, 0.00009. As can be seen, there is a sudden drop in the value of the fourth Eigen value relative to the third one. This is a reasonable indicator that the system is third order. The 3x3 Hakel and the shifted Hankel matrices are:

$$\text{Ha} = \begin{bmatrix} -0.18 & 10.2 & -7.04 \\ 10.2 & -7.04 & -041.3 \\ -7.04 & -041.3 & 223.1 \end{bmatrix}$$

$$\widetilde{\text{Ha}} = \begin{bmatrix} 10.2 & -7.04 & -41.3 \\ -7.04 & -41.3 & 223.1 \\ -41.3 & 223.1 & -651.6 \end{bmatrix} \quad (57)$$

The resulting state space model is (58):

$$\begin{bmatrix} \dot{x}_1 \\ \dot{x}_2 \\ \dot{x}_3 \end{bmatrix} = \begin{bmatrix} 0 & 1 & 0 \\ 0 & 0 & 1 \\ -.013 & -6.96 & -4.21 \end{bmatrix} \begin{bmatrix} x_1 \\ x_2 \\ x_3 \end{bmatrix} + \begin{bmatrix} -.18 \\ 10.2 \\ -7.04 \end{bmatrix} \cdot V_a \quad (58)$$

$$[x_1] = \begin{bmatrix} 1 & 0 & 0 \end{bmatrix} \begin{bmatrix} x_1 \\ x_2 \\ x_3 \end{bmatrix}$$

Directly computing the transfer function from the state equation yields (59):

$$H(S) = \frac{1}{5.92} \cdot \frac{-.18 \cdot S^2 + 9.45 \cdot S + 34.67}{S^3 + 4.51 \cdot S^2 + 6.53 \cdot S + .013} \quad (59)$$

The transfer function is approximated to yield the position and velocity transfer functions of the trainer (60):

$$H_p(S) = \frac{1}{5.92} \cdot \frac{9.45 \cdot S + 34.67}{S \cdot (S^2 + 4.51 \cdot S + 6.53)} \quad (60)$$

$$H_v(S) = \frac{1}{5.92} \cdot \frac{9.45 \cdot S + 34.67}{S^2 + 4.51 \cdot S + 6.53} \quad (61)$$

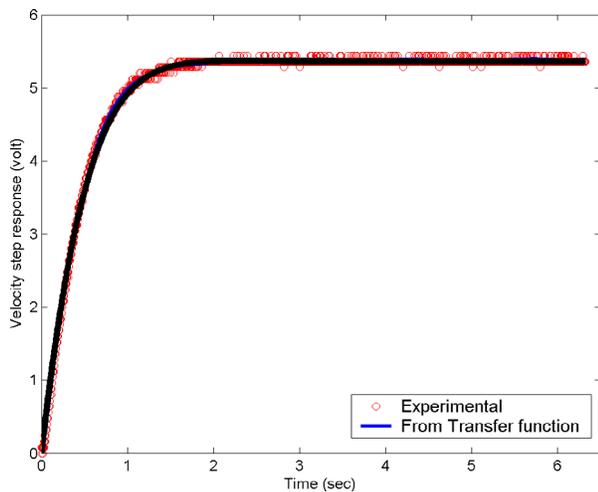

Figure-35: Experimental speed step response and the one predicted from the estimated transfer function-Markov approach

The step response of the estimated velocity transfer function is shown in figure-35 along with the experimental one. As can be seen, the transfer function provides good fit of the experimental response.

It is interesting to notice that the velocity transfer function of the trainer is not only second order, it also has a zero that is most probably caused by the power amplifier. This observation is corroborated by the extended model results form the pseudo-based inversion method.

Further verification of the motor model (60, 61) is carried-out using the feedback servo trainer 33-110 onboard test signals. In addition to a step input, the trainer provides sinusoidal, triangular and square wave test signals. The magnitude of the test waves is set to about 6.4 volt and their frequency to about 0.1 HZ. The same connection in figure-23 is used with the exception of the motor being fed from the respective test signal. The motor is excited by each of the reference waveforms and the speed response is recorded. This reference is then fed to the estimated motor's velocity transfer function (61) and a reconstruction of the response is obtained. The used sinusoidal reference input along with the experimental and constructed motor speed response are shown in figure-36. The results for the triangular and square inputs are shown in figures-37 and 38 respectively. As can be seen, the estimated transfer function provided good fit for all the cases.

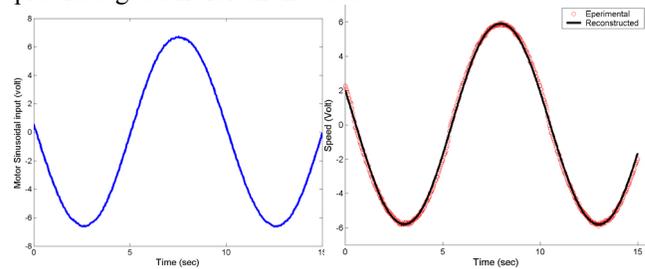

Figure-36: Reference, experimental and constructed speed response, sinusoidal reference.

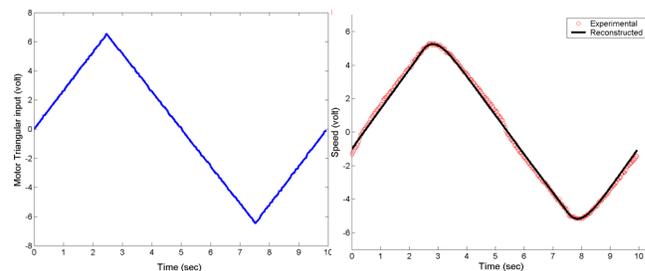

Figure-37: Reference, experimental and constructed speed response, triangular reference.

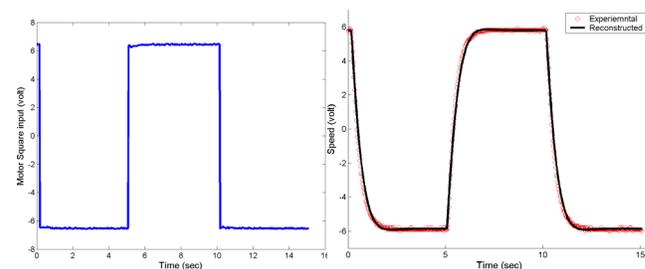

Figure-38: Reference, experimental and constructed speed response, square pulse reference.

## VII. STUDENT FEEDBACK

After the pseudo-based motor identification experiment was performed by the students, they were asked to write a report. About a week after the report was submitted the experiment was assessed directly and indirectly.

Direct assessment involves:
- Ability of the student to finish the experiment in one lab session (2:45 hrs)
- Adequacy of the documentation observed in terms of the assistance the students requested from the laboratory supervisor
- The quality of the experiment report which each student had to hand-in individually
- The observed attitude of the students towards the experiment.

The following was observed:
- The experiment was in general well-received by the students. They expressed interest not only in the SS identification procedure, but also in the supporting tools. Some were having their capstone design project and mentioned that they used some of these tools in their projects
- All students were able to finish the experiment in one laboratory session using the handouts with minimal dependence on the laboratory supervisor
- Most of the students wrote good reports. A sizable number went beyond what is expected, and wrote excellent thorough reports. The report mark distribution for 52 students is shown in figure-39

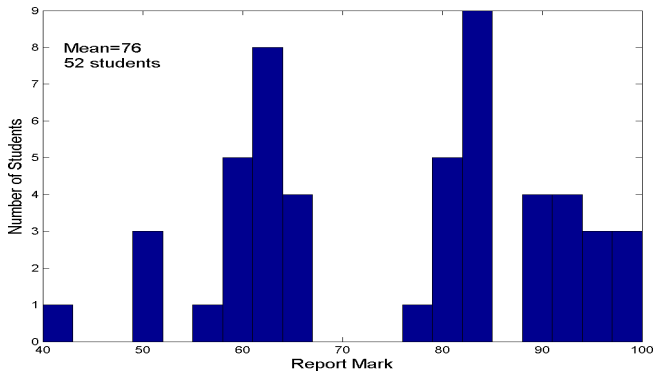

Figure-39: Mark distribution of the experiment reports

For the indirect assessment, a questionnaire was distributed to the students in one laboratory section in order to get their feedback about the experiment. Below are some of the questions the questionnaire contains:

1. The objective of the experiment is clear
2. The introduction is helpful in understanding and performing the experiment
3. The level of experiment is suitable for an undergraduate course in control
4. The equipment in the lab are enough to effectively perform the experiment
5. The experiment is strongly related to the control theory covered in class
6. The experiment is practical and will be helpful in the future
7. The experiment introduced me to useful advanced mathematical tools.

The assessment questions are rated on a scale from -5 to +5. The student were guided through this scale by the five options: SA which corresponds to strongly agree, A for agree, N for neutral, D and SD for disagree and strongly disagree (figure-40). The students were also asked to record any comment they have. Table-4 shows the results from the eighteen students who filled the questionnaire. The experiment seems to have been well-received by them.

## VIII. CONCLUSION

Two student-friendly undergraduate procedures are suggested for experimentally identifying the state space model of a DC motor in a basically equipped control laboratory. The first procedure is based on the direct pseudo-inversion of the state space model. It requires minimal exposure to state space theory. The second one is based on the Markov approach for state space realizations. It is much more robust and accurate than the first one. However, it requires a senior undergraduate or a 1st graduate level exposure to state space theory. Both procedures place no restrictions on the motor's model. Both approaches use the laboratory friendly and easy to implement step input (switch on the motor and observe the response). Data collection is easily performed using standard laboratory equipment. Data processing can be carried-out using Matlab and Excel. The experiments have good informative component that introduces an undergraduate student to advanced mathematical tools with direct tangible laboratory outcomes. The suggested experimental procedures are adaptable for use in advanced and industrial control laboratories. They can also be used in laboratories other than control, e.g. advanced circuit laboratories.

| -5 | -4 | -3 | -2 | -1 | 0 | +1 | +2 | +3 | +4 | +5 |
|---|---|---|---|---|---|---|---|---|---|---|
| Strongly disagree | Disagree | | | Neutral | | | Agree | | Strongly Agree | |

Figure-40: scale used in evaluating the questions

Table-4: The results from the student evaluation

| Question # | SA | A | N | D | SD | mean |
|---|---|---|---|---|---|---|
| 1 | 10 | 7 | 1 | 0 | 0 | 3.56 |
| 2 | 5 | 6 | 5 | 2 | 0 | 1.94 |
| 3 | 9 | 5 | 3 | 1 | 0 | 2.95 |
| 4 | 13 | 1 | 2 | 2 | 0 | 3.33 |
| 5 | 6 | 7 | 4 | 1 | 0 | 2.39 |
| 6 | 7 | 5 | 6 | 0 | 0 | 2.83 |
| 7 | 7 | 7 | 3 | 1 | 0 | 2.67 |
| Overall average | 46% | 30% | 19% | 5% | 0% | 2.81 |

**ACKNOWLEDGMENT**
The authors gratefully acknowledge the assistance of king Fahad University of Petroleum and Minerals.

**REFERENCES AND FOOTNOTES**
[1] G. W. Younkin, Industrial servo control systems: fundamentals and applications. New York: Dekker, 2003.
[2] R. C. Dorf and R. H. Bishop, Modern control systems. Essex: Pearson, 2017.
[3] P. Clement, "A note on third-order linear systems," IRE Transactions on Automatic Control, vol. 5, no. 2, pp. 151–151, 1960.
[4] R. Monzingo, "On approximating the step response of a third-order linear system by a second-order linear system," IEEE Transactions on Automatic Control, vol. 13, no. 6, pp. 739–739, 1968.
[5] Sendrescu, D. , "DC Motor Identification Based on Distributions Method," Annals of the University of Craiova, no. 36, vol.9, pp. 41-49, 2012.

[6] AL-qassar, R., Othman, M., "Experimental Determination of Electrical and Mechanical Parameters of DC Motor Using Genetic Elman Neural Network", Jour. of Eng. Sci. and Tech., No. 2, Vol.3, pp. 190 – 196, 2008.

[7] Mohammed S. Z. Salah, "Parameters Identification of A Permanent Magnet DC Motor", M.Sc, Electrical engineering, IUG, 2009

[8] S. Udomsuk, K.-L. Areerak, K.-N. Areerak, and A. Srikaew, "Parameters identification of separately excited DC motor using adaptive tabu search technique," 2010 International Conference on Advances in Energy Engineering, 2010.

[9] W. Wu, "DC motor identification using speed step responses," Proceedings of the 2010 American Control Conference, 2010.

[10] P. M. Makila, "State space identification of stable systems," International Journal of Control, vol. 72, no. 3, pp. 193–205, 1999.

[11] M. Viberg, "Subspace-based methods for the identification of linear time-invariant systems," Automatica, vol. 31, no. 12, pp. 1835–1851, 1995.

[12] J. Basilio and M. Moreira, "State–Space Parameter Identification in a Second Control Laboratory," IEEE Transactions on Education, vol. 47, no. 2, pp. 204–210, 2004.

[13] H. Garnier, "Direct continuous-time approaches to system identification. Overview and benefits for practical applications," European Journal of Control, vol. 24, pp. 50–62, 2015.

[14] A. Griewank and A. Walther, Evaluating derivatives principles and techniques of algorithmic differentiation. Philadelphia: SIAM, 2008.

[15] S. Stergiopoulos, Advanced signal processing handbook: theory and implementation for radar, sonar, and medical imaging real time systems. Boca Raton: CRC Press, 2019.

[16] V. Verma, C. Baird, and V. Aatre, "Pulse-width-modulated speed control of d.c. motors," Journal of the Franklin Institute, vol. 297, no. 2, pp. 89–101, 1974.

[17] https://www.parkermotion.com/manuals/6270/6270_3.pdf

[18] C. Chen, "Linear System Theory and Design", Oxford University Press, 2013.

[19] Ahmad A. Masoud, Mohammad Abu-Ali, Ali Al-Shaikhi, "A Novel and Effective Procedure for an Undergraduate First Control Laboratory Education in State Space-based Motor Identification", 2018 Annual American Control Conference (ACC) June 27–29, 2018. Wisconsin Center, Milwaukee, USA, pp.4843-4848

[20] M. Abramowitz, and I. Stegun. , "Handbook of Mathematical Functions with Formulas, Graphs, and Mathematical Tables", Dover Publications, New York, (1964)

[21] A. V. Oppenheim, R. W. Schafer, "Discrete-Time Signal Processing", Prentice-Hall International Inc., 1989

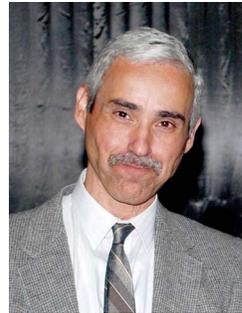

**Ahmad A. Masoud** received his B.Sc. degree in electrical engineering with a major in power systems and a minor in communication systems from Yarmouk University, Irbid, Jordan, in 1985, and his M.Sc. in signal processing and Ph.D. in robotics and autonomous systems both from the electrical engineering department at Queen's University, Kingston, Ontario, Canada, in 1989 and 1995, respectively. He worked as a researcher with the Electrical Engineering Department, Jordan University of Science and Technology, Irbid, from 1985 to 1987. He was also a part-time assistant professor and a research fellow with the Electrical Engineering Department, Royal Military College of Canada, Kingston, from 1996 to 1998. During that time, he carried out research in digital signal processing-based demodulator design for high density, multiuser satellite systems and taught courses in robotics and control systems. He is currently an assistant professor with the Electrical Engineering Department, King Fahd University of Petroleum and Minerals, Dhahran, Saudi Arabia. His current interests include navigation and motion planning, robotics, constrained motion control, intelligent control, and DSP applications in machine vision and communication

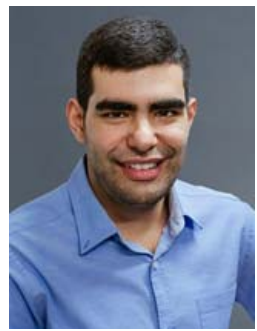

**Mohammad Abu-Ali** received the B.Sc. degree (with First Honors) in Electrical Engineering from King Fahd University of Petroleum and Minerals, Dhahran, Saudi Arabia, in 2017. He is currently pursuing his master's degree in Power Engineering at the Technical University of Munich, Munich, Germany. His research interests include control and optimization of mechatronic drives and power electronics.

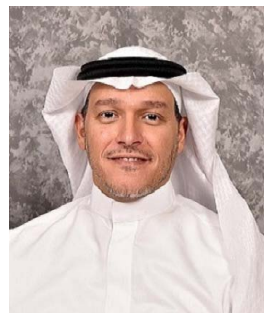

**Ali Al-Shaikhi** is an Associate Professor and currently the Dean of Engineering College at KFUPM. He received the B.Sc. and M.Sc. degrees from the King Fahd University of Petroleum and Minerals (KFUPM), Dhahran, Saudi Arabia, in 1997 and 2001, respectively, and the Ph.D. degree from Dalhousie University, Halifax, Canada, in 2008, all in electrical engineering. He has been the Chairman of the Electrical Engineering Department, KFUPM, since 2011 and the Director of the Center for Energy and Geo-Processing (CEGP) since 2013. He serves as a member of the following committees: Academic Leadership Center (ALC) at Ministry of Education (MOE), Electricity Dispute Resolution Committee at ECRA, Lighting Team at SASO, and External Advisory Board of Electrical Engineering Department at Umm Al-Qura University. He received the following scholarships: the faculty of graduate study at Dalhousie University Scholarship and Bruce and Dorothy Rosetti Engineering Research Scholarship. He also received a Travel Grant Award of the 5th Annual Conference on Communication Networks and Services Research, in 2007. He was awarded KFUPM Scholarship to pursue his Ph.D. degree overseas. He has several publications in ISI journal, flagship conferences, and has some patents, as well. His research interests lie in the areas of wireless communication, communication theory, digital communications, signal processing, and computer networks.